\documentstyle[sprocl,epsfig]{article}

\def\bm#1{\hbox{\boldmath $#1$}}

\def\be{\begin{equation}}
\def\ee{\end{equation}}
\def\bea{\begin{eqnarray}}
\def\eea{\end{eqnarray}}
\begin{document}

\pagestyle{empty}

\title{MULTI-PHASE HYDRODYNAMICS AND X-RAY CLUSTERS FORMATION}

\author{Romain TEYSSIER and Jean-Pierre CHI{\`E}ZE}

\address{CEA, DSM/DAPNIA/Service d'Astrophysique,
\\CE Saclay, F-91191 Gif-sur-Yvette Cedex, France
\\E-mail: rteyssie@discovery.saclay.cea.fr}

\author{Jean-Michel ALIMI}

\address{Laboratoire d'Astrophysique Extragalactique et de Cosmologie,
\\CNRS URA 173, Observatoire de Paris-Meudon, 92195 Meudon, France
\\E-mail: alimi@gin.obspm.fr}


\maketitle\abstracts{ We investigate the role of radiative
cooling within the core of large X-ray clusters using multi-phase
hydrodynamics. We developed for that purpose a spherically
symmetric hydrodynamical code, coupled to a ``fluid model" that
describes accurately the dark matter component. Cooling is included
using a self-consistent multi-phase approach, leading to ``cooled
gas" mass deposition throughout the flow. We simulate the collapse
and the subsequent evolution of a Coma-like X-ray cluster, avoiding
the well-known ``cooling catastrophe". The total mass profile of
our simulated cluster is very similar to the ``universal" profile
proposed by Navarro, Frenk \& White (1995)~\cite{navarro95}. More
interestingly, we also obtain a quasi-isothermal temperature
profile, which is a direct consequence of multi-phase cooling
within such a potential well.}

\section{Introduction}

Active cooling strongly regulates the evolution of the dynamically
relaxed central regions of many galaxy clusters, as inferred from
X-ray observations~\cite{edge92}. The analysis of X-ray spectra of
these regions often exhibit an excess of low-energy photons,
relative to a single-temperature spectrum. Recently, high EUV
brightness excess have been detected by EUVE pointed observations
of the Virgo and Coma clusters~\cite{lieu95}. Such observations
strongly suggest the presence of cold gas at a temperature well
below the average X-ray temperature of the intra cluster gas. This
may be an evidence for the presence of cold parcels of gas
immersed in the hot, pervasive X-ray emitting gas.

Gas  cooling is a complex, unstable phenomenon. As a matter of
fact, the gas is subject to the so-called {\it cooling instability}
by which small temperature differences are amplified, leading to a
clumpy structure on small scales. A long lived, tenuous hot phase
can be maintained as the bulk of the mass cools down and eventually
condense in cold clouds and, possibly, stars. This description of
the thermal history of the gas has been investigated by numerous
authors, and is known has the ``multi-phase cooling flow"
formalism~\cite{nulsen86,lioure90,waxman95,thomas96}.

In this paper, we intend to investigate multi-phase cooling using a
fully hydrodynamical approach. We therefore proceed a step further
than the usual stationary cooling flow approach. We describe the
dynamical evolution of both gas and dark matter
components~\cite{chieze97a}, starting from cosmologically relevant
initial conditions, together with the thermodynamical evolution of
the multi-phase medium, which ultimately leads to the condensation
of cold clouds. We shall restrict ourselves to very high
resolution, spherically symmetric simulations, since we are
interested in the most central regions of X-ray clusters, where
strong cooling takes place.

\section{Physical and Numerical Methods}

Nulsen (1986) derived the equations describing the evolution of the
density distribution in a multi-phase medium. These multi-phase
equations describe the flow at intermediate scales, in between
microscopic scales, where atomic processes contribute to gas
cooling, and macroscopic scales, relevant for the overall
hydrodynamical evolution. Integrating the multi-phase equations
over the density distribution, one obtains useful {\it modified
hydrodynamical equations}. First, the continuity equation writes

\be
\frac{1}{\bar \rho}\frac{d\bar\rho}{dt}+\bm \nabla \cdot \bm u =
-\beta(x,t)
\ee

\noindent
where $\beta$ is {\bf the mass deposition rate} and $\bar \rho$ is
the {\it mean} density in the macroscopic fluid element. Second,
the energy equation writes

\be
\frac{1}{P} \frac{dP}{dt} +\frac{5}{3}\bm \nabla \cdot \bm u =
\frac{2}{3}\lambda(x,t)\frac{{\bar n}^2 \Lambda(\bar T)}{P}
\ee

\noindent
where $P$ is the gas pressure, $\Lambda(\bar T)$ the cooling
function evaluated at the {\it mean} temperature, and
$\lambda(x,t)$ the {\bf cooling enhancement factor}, due to the
presence of an underlying density distribution. Finally, the Euler
equation writes

\be
\bar \rho \frac{d\bm u}{dt} = -\bf \nabla P - \bar \rho \bf \nabla \Phi
\ee

\noindent
The usual single phase hydrodynamical equations are recovered with
$\lambda=1$ and $\beta=0$. However, in the general multi-phase
case, the functions $\beta(x,t)$ and $\lambda(x,t)$ have to be
evaluated through a detailed treatment of the multi-phase
distribution. This has been done so far only in the case of
stationary cooling flows~\cite{waxman95,thomas96}. It has been
however demonstrated that, given any reasonable initial
multi-phase medium, a universal high-density cooling tail rapidly
develops in the mass spectrum, leading to self-similar
solutions~\cite{nulsen86}. In this case, the cooling enhancement
factor $\lambda$ is constant all over the flow, and the mass
deposition rate $\beta$ is exactly proportional to the {\it mean}
cooling rate in the macroscopic fluid element. In this paper, we
neglect the short relaxation phase anterior to this asymptotic
cooling flow regime. We consider only the self-similar solutions
found by Nulsen (1986), parameterized by a single parameter $\nu$.
Details on our numerical treatment will be found
elsewhere~\cite{chieze97b}.

\begin{figure}[t]
\hskip 1.2cm
\psfig{figure=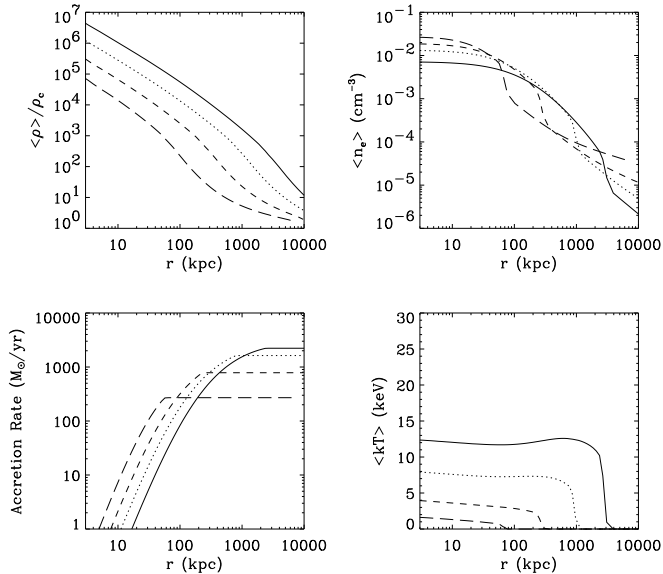,height=8cm}
\caption{
Simulated profiles of a Coma-like cluster: enclosed overdensity,
{\it mean} electron density, accretion rate and {\it mean} gas
temperature at different redshifts (z=2.4, 1.2, 0.5, 0). Note that
the X-ray temperature is here only 70\% of the {\it mean}
temperature.
\label{fig1}}
\end{figure}

\section{Results and Discussion}

We simulate the formation of a large X-ray cluster, similar to the
Coma cluster ($M_{tot}\simeq 1\times10^{15}$, $\bar T\simeq 10^8$).
We plot in figure (\ref{fig1}) the profiles we obtained at
different redshifts. The dark matter density profile is similar to
the Navarro, Frenk et White (1995) universal profile, with
$\rho\propto r^{-1}$ in the center. We found that only large values
of the multi-phase parameter are allowed ($\nu \ge 3$), in order to
avoid a cooling catastrophe. This roughly corresponds to the
limiting case $\nu = +\infty$, which allows indeed a strict
conservation of the gas mean entropy, as fluid elements sink
towards the center. It turns out to be the only stable case. This
case corresponds to a wide temperature distribution within the
intracluster medium, with small parcels of gas of arbitrary high
temperature.

The X-ray emitting gas density profile show a typical core-halo
structure, the core radius corresponding to the cooling radius of
the cluster. Note that the central density decreases in time, due
to continuous mass deposition in this strongly cooling region. This
deposition process regulates the mass infall, and the central
density remains at a value corresponding to $t_{cool} \simeq t_H$.
The accretion rate steadily increases with radius as $\dot M\propto
r^{2}$, and reaches 2000 $M_{\odot}/yr$ in the outer regions, which
is observed in strong cooling flow clusters. More interestingly,
the mean temperature profile is nearly isothermal. This property is
due, first, to the total mass profile ($\rho \propto r^{-1}$),
which determines the gravitational potential, and, second, to the
multi-phase model we considered ($\nu = +\infty$). We found also
that the inner temperature profile is highly sensitive to the mass
profile. This lead us to conclude that, within our multi-phase
model, the total density profile $\rho \propto r^{-1}$ is the only
one leading to an isothermal gas distribution.

The multi-phase model ($\nu=+\infty$) we use here might be
considered as unrealistic, since it involves a very wide
temperature distribution in the intracluster gas. We believe on the
contrary that this model mimics the continuous reinjection of high
entropy gas by supernovae driven winds, a very important aspect of
clusters thermodynamics. A more realistic treatment of multi-phase
cooling is however necessary, and we intend to include within our
hydrodynamical code a complete (multi-group) treatment of the mass
deposition.

\end{document}